\documentclass[english,aps,preprint]{revtex4}
\usepackage[]{fontenc}
\usepackage[latin9]{inputenc}
\usepackage{graphicx}
\usepackage{amssymb}
\usepackage{babel}

\begin{document}

\title{Power law behavior for the zigzag transition in a Yukawa cluster}

\author{T. E. Sheridan}

\email{t-sheridan@onu.edu}

\author{Andrew L. Magyar}

\altaffiliation{Present address: Department of Physics, Penn State University, University Park, PA 16802}

\affiliation{Department of Physics and Astronomy, Ohio Northern University, Ada,
OH 45810}

\begin{abstract}
We provide direct experimental evidence that the one-dimensional (1D)
to two-dimensional (2D) zigzag transition in a Yukawa cluster exhibits
power law behavior. Configurations of a six-particle dusty (complex)
plasma confined in a biharmonic potential well are characterized as
the well anisotropy is reduced. When the anisotropy is large the particles
are in a 1D straight line configuration. As the anisotropy is decreased
the cluster undergoes a zigzag transition to a 2D configuration. The
measured dependence of cluster width on anisotropy is well described
by a power law. A second transition from the zigzag to an elliptical
configuration is also observed. The results are in very good agreement
with a model for particles interacting through a Yukawa potential.
\end{abstract}


\maketitle
Plasma is a quasi-neutral gas that contains electrons, ions and (often)
neutral atoms and which exhibits collective behavior \cite{chen}.
A dusty (complex) plasma is a systems of interacting microscopic dust
particles immersed in a standard electron-ion plasma. For typical
laboratory conditions, dust particles acquire a net charge $q<0$
from the electron and ion currents. The Coulomb interaction between
the charged dust particles is shielded by the response of the free
charge in the plasma. In the plane of a two-dimensional cluster, the
inter-particle potential is well approximated by a Yukawa (i.e., Debye
or shielded Coulomb) potential \cite{lam}\begin{equation}
V\left(r\right)=\frac{1}{4\pi\epsilon_{0}}\,\frac{q}{r}e^{-r/\lambda_{D}},\label{eq:Vr}\end{equation}
where $\lambda_{D}$ is the Debye screening length. An unshielded
Coulomb interaction is recovered as $\lambda_{D}\rightarrow\infty$.
Dusty plasma is an exceptional experimental system for examining the
static and dynamic properties of Yukawa clusters--small clusters of
particles interacting through a Yukawa potential.

In laboratory experiments the dusty plasma is usually confined, typically
through a combination of gravitational and electrical forces. Monodisperse
microspheres may then form strongly-coupled one-dimensional (1D) \cite{mis,hom,liu},
two-dimensional (2D) \cite{juan,tes3,tesj} or even three-dimensional
systems \cite{arp}. The general 2D potential well is biharmonic for
small displacements from the potential energy minimum \cite{tes4,mel1,can}.
A dusty plasma confined in a 2D biharmonic well may be either in a
1D straight line configuration, or a 2D configuration, where the transition
from 1D to 2D states is via the zigzag instability \cite{mel1,sch,can,apo,tesps}.

A cold 2D dusty plasma can be modeled \cite{can,tes8,tes4} by assuming
$n$ identical particles with mass $m$ and charge $q$, where the
$i$th particle is at $\left(x_{i},y_{i}\right)$. The potential energy
$U$ of a configuration $\left\{ x_{i},y_{i}\right\} $ is given by
the sum of the particles\textquoteright{} potential energy with respect
to the biharmonic well plus the sum of the potential energy for each
pair-wise interaction,\begin{equation}
U=\sum_{i=1}^{n}\frac{1}{2}m\left(\omega_{x}^{2}x_{i}^{2}+\omega_{y}^{2}y_{i}^{2}\right)+\sum_{i=1}^{n}\sum_{j>i}^{n}\frac{1}{4\pi\epsilon_{0}}\,\frac{q^{2}}{r_{ij}}e^{-r_{ij}/\lambda_{D}},\label{eq:U}\end{equation}
where $r_{ij}=\sqrt{\left(x_{i}-x_{j}\right)^{2}+\left(y_{i}-y_{j}\right)^{2}}$
is the separation between particles $i$ and $j$, and $\omega_{x}$
and $\omega_{y}$ are the single-particle oscillation frequencies
in the $x$ and $y$ directions, respectively. The equilibrium configuration
minimizes $U$.

We nondimensionalize Eq. (\ref{eq:U}) using the variables $\xi_{i}=x_{i}/r_{0}$,
$\eta_{i}=y_{i}/r_{0}$, $\rho_{ij}=r_{ij}/r_{0}$ and parameters\begin{equation}
\alpha^{2}=\frac{\omega_{y}^{2}}{\omega_{x}^{2}},\;\kappa=\frac{r_{0}}{\lambda_{D}}.\label{eq:a2k}\end{equation}
Here the characteristic length \begin{equation}
r_{0}^{}=\left(\frac{2}{m\omega_{x}^{2}}\,\frac{q^{2}}{4\pi\epsilon_{0}}\right)^{1/3}\label{eq:r0}\end{equation}
is defined using the longitudinal frequency $\omega_{x}$. The dimensionless
potential energy is then\begin{equation}
\frac{U}{U_{0}}=\sum_{i=1}^{n}\left(\xi_{i}^{2}+\alpha^{2}\eta_{i}^{2}\right)+\sum_{i=1}^{n}\sum_{j>i}^{n}\frac{e^{-\kappa\rho_{ij}}}{\rho_{ij}},\label{eq:UU0}\end{equation}
where $U_{0}$ is the characteristic potential energy. Equilibrium
configurations $\left\{ \eta_{i},\xi_{i}\right\} $ of the model {[}Eq.
(\ref{eq:UU0})] depend on three parameters: $n$, $\kappa$ and $\alpha^{2}$.
Here $\kappa$ is the Debye shielding parameter and $\kappa=0$ corresponds
to an unshielded Coulomb interaction. The well anisotropy parameter
is $\alpha^{2}$. We assume $\omega_{y}\ge\omega_{x}$ or $\alpha^{2}\ge1$.
Straight line configurations $y_{i}=\eta_{i}=0$ then lie in the $x$
(longitudinal) direction and are independent of $\alpha^{2}$. An
equilibrium configuration $\left\{ \xi_{i},\eta_{i}\right\} $ can
be compared to an experimentally measured configuration $\left\{ x_{i},y_{i}\right\} $
by multiplying each coordinate by $r_{0}$. 

Within the context of this model {[}Eq. (\ref{eq:UU0})], abrupt 1D
to 2D transitions \cite{tes8,can} are predicted to occur for critical
values of the three parameters $n$, $\kappa$ and $\alpha^{2}$ due
to a zigzag instability. This has been confirmed experimentally as
$n$ was varied by Melzer \cite{mel1} and Sheridan and Wells \cite{tes8}.
Melzer \cite{mel1} also observed a zigzag transition as the neutral
gas pressure was changed, which presumably changed the plasma density
and therefore both $\kappa$ and $\alpha^{2}$. Sheridan and Wells
\cite{tes8} confirmed that the zigzag transition caused by increasing
$n$ in small clusters behaves as a continuous phase transition between
1D and 2D states, where the cluster width near the transition obeys
a power law. Using the model of Eq. (\ref{eq:UU0}), they also predicted
\cite{tes8} that the zigzag transition in small clusters exhibits
power law behavior as either $\kappa$ or $\alpha^{2}$ is varied.
Piacente, et al. \cite{pia} reached a similar conclusion from a computational
study of unbounded Yukawa systems ($n\rightarrow\infty$) with a transverse
parabolic well and longitudinal periodic boundaries. Structural transitions
due to changes in $\kappa$ or $\alpha^{2}$ have not been characterized
experimentally.

In this work we investigate the transitions a Yukawa cluster (i.e.,
a small dusty plasma) undergoes as the well anisotropy $\alpha^{2}$
is varied. We use a novel experimental method in which the physical
geometry of the confining potential well is changed without disrupting
the plasma discharge. The discharge parameters are nearly constant,
so that the Debye shielding parameter $\kappa$ is also nearly constant.
The well anisotropy is accurately determined using measured center-of-mass
frequencies for the dusty plasma. As predicted \cite{tes8}, the transverse
cluster width $y_{rms}$ exhibits power law behavior vs $\alpha^{2}$
following the zigzag transition. The Debye shielding parameter is
found by comparison to the model, and Debye shielding is shown to
be important.

\begin{figure}
\includegraphics[width=3.25in]{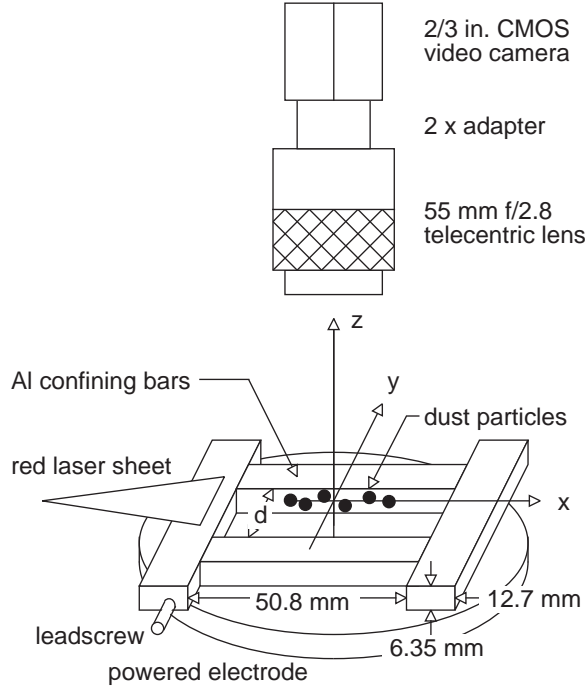}

\caption{\label{fig:Schem}Schematic of the experimental geometry (not to scale).
Six dust particles are confined in the biharmonic potential well created
by a rectangular depression placed on the powered electrode in an
rf discharge. The confining well anisotropy is varied by changing
the width $d$ while the plasma is on, and without losing the particles,
using a lead screw.}

\end{figure}

The dusty plasma was studied in the DONUT (Dusty Ohio Northern University
experimenT) apparatus (Fig. \ref{fig:Schem}) \cite{tes2,tes3,tes4,tes6,tes8,tesj}.
Argon gas was leaked into the vacuum chamber and the pressure was
stabilized at 15.4 mtorr. At this pressure normal modes of the dust
clusters are under damped \cite{tes2}. Inside the chamber a capacitively-coupled
rf electrode is used both to sustain the discharge and to levitate
the dust particles. The rf frequency was 13.56 MHz, the forward rf
power was $\sim7$ W and the dc self-bias on the electrode was $-90\;{\rm V}$.

The confining well (Fig. \ref{fig:Schem}) is produced by a rectangular
aperture placed on the 89-mm diameter aluminum electrode, creating
a biharmonic well in the horizontal ($x$-$y$) plane \cite{hom}.
The aperture is made of four aluminum bars, each with a cross section
of 6.35 mm $\times$ 12.7 mm. The depth of the trap is 6.35 mm, the
length is fixed at 50.8 mm, and the width $d$ is adjustable from
$\approx0$ to 38 mm. Here $d$ can be adjusted while the plasma is
on, and without losing the dust particles, by mean of a lead screw
which is rotated from outside the vacuum. The lead screw is made of
joined left-handed and right-handed threaded rods {[}unified national
coarse (UNC) thread size 6-32] that move the two confining bars simultaneously
closer together or farther apart a distance of 1/16 inch ($\approx1.59\;{\rm mm}$)
per revolution, thus keeping the minimum of the potential well centered
with respect to the disk electrode. This arrangement allows us to
continuously vary the anisotropy parameter $\alpha^{2}$ while maintaining
nearly the same plasma discharge parameters. 

The dust particles used were melamine formaldehyde resin spheres with
a nominal diameter of $9.62\pm0.09\;\mu{\rm m}$. As noted previously
\cite{tes2}, we believe that the particle diameter is actually $8.94\;\mu{\rm m}$,
giving $m=5.65\times10^{-13}\;{\rm kg}$. Confined dust particles
are illuminated by a red diode laser sheet. A top-view camera mounted
about 30 cm above the electrode is used to acquire dust particle position
data. For this experiment, the resolution of the camera and associated
optics was 16.47 $\mu$m/pixel. Sequences of 2048 frames were recorded
at 27.9 frames/s. A side-view camera was used to confirm that there
were no out-of-plane particles. We acquired 20 data sets with the
same $n=6$ particles for trap widths $d=12.8$ to 36.6 mm.

\begin{figure}
\includegraphics[width=3in]{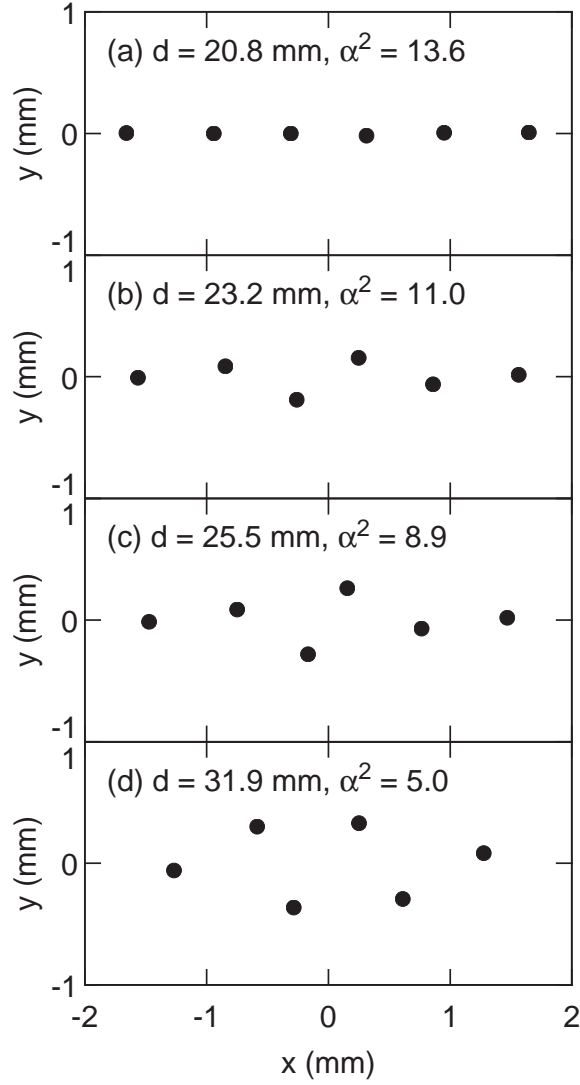}

\caption{\label{fig:config}Measured particle configurations vs increasing
trap width $d$, or equivalently, decreasing well anisotropy $\alpha^{2}$.
(a) Straight line configuration, (b) and (c) zigzag configurations
and (d) elliptical configuration. Values of the anisotropy parameter
$\alpha^{2}$ were determined from measured c.m. frequencies.}

\end{figure}

Representative measured configurations are shown in Fig. \ref{fig:config}
for increasing $d$. As seen in Fig. \ref{fig:config}(a), the particles
are in a straight line configuration when $d$ is small, or equivalently,
$\alpha^{2}$ is large. As we increase $d$ the cluster goes through
two structural transitions, straight line to zigzag {[}Figs. \ref{fig:config}(b)
and (c)], and then zigzag to elliptical {[}Fig. \ref{fig:config}(d)].
The transition to the zigzag state occurs at $d\approx23\;{\rm mm}$
and the transition to the elliptical state occurs at $d\approx30\;{\rm mm}$.
In the elliptical state all six particles lie on the convex hull of
the cluster.

To analyze the experimental data, we first find the dependence of
the anisotropy parameter $\alpha^{2}$ on $d$. This is done by measuring
the frequencies $\omega_{x}$ and $\omega_{y}$ of the c.m. modes
from the particles' Brownian motions \cite{tes3,tes4,tes8}. The power
spectra of the time histories of the c.m. coordinates, $x_{cm}=\left(1/n\right)\sum x_{i}$
and $y_{cm}=\left(1/n\right)\sum y_{i}$, were fitted with the expression
for a driven damped harmonic oscillator to determine the corresponding
normal mode frequencies $\omega_{x}$ and $\omega_{y}$. For each
$d$, the dust particle configuration was characterized by the root-mean-squared
values $x_{rms}$ and $y_{rms}$ averaged over all frames, where\begin{equation}
x_{rms}^{2}=\frac{1}{n}\sum_{i=1}^{n}\left(x_{i}-x_{cm}\right)^{2},\; y_{rms}^{2}=\frac{1}{n}\sum_{i=1}^{n}\left(y_{i}-y_{cm}\right)^{2}.\label{eq:xrms}\end{equation}
Here $x_{rms}$ represents the longitudinal cluster size (i.e., its
length), while $y_{rms}$ is the transverse cluster width. In particular,
$y_{rms}$ can be used as an order parameter for the zigzag transition
since $y_{rms}=0$ in the 1D straight line configuration and $y_{rms}>0$
for the 2D zigzag configuration \cite{tes8}.

\begin{figure}
\includegraphics[width=3.25in]{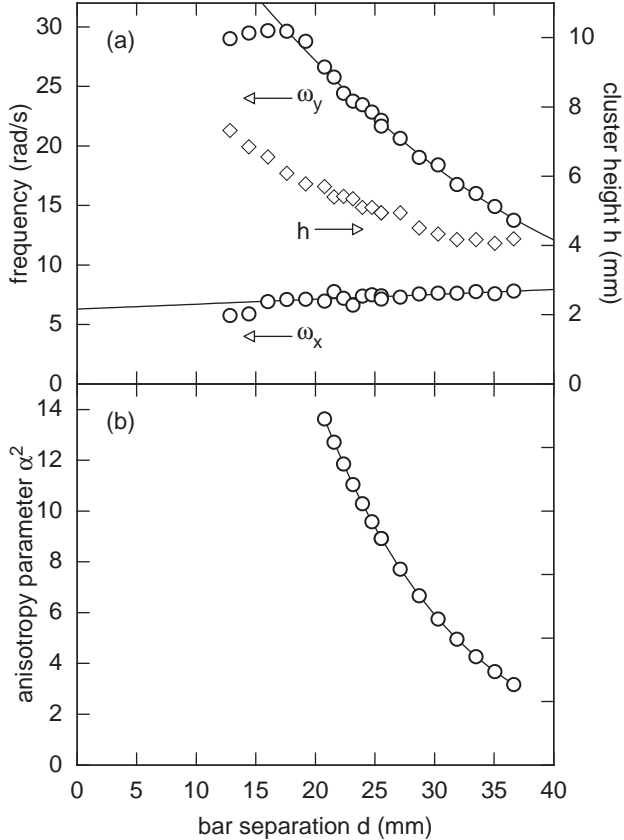}

\caption{\label{fig:wxwya2}(a) Dependence of measured c.m. frequencies $\omega_{x}$
and $\omega_{y}$ and the height $h$ of the cluster above the top
of the well on the width $d$. Here $\omega_{x}$ is fitted with a
linear function and $\omega_{y}$ with a quadratic curve. (b) Dependence
of the smoothed anisotropy parameter $\alpha^{2}=\omega_{y}^{2}/\omega_{x}^{2}$
on $d$. }

\end{figure}

The measured c.m. frequencies $\omega_{x}$ and $\omega_{y}$ are
plotted vs $d$ in Fig. \ref{fig:wxwya2}(a). Here $\omega_{x}$ (the
frequency associated with the fixed 50.8 mm separation) shows a weak
linear increase with $d$, while $\omega_{y}$ is roughly constant
for $d\lesssim20$ mm and then decreases as $d$ increases. That is,
for these experimental conditions decreasing $d$ below $\approx20$
mm does not increase $\omega_{y}$ because the sheath edge can no
longer conform to the rectangular depression, but rather, is pushed
upward \cite{mel1}. This explanation is given further weight by measurements
of the cluster height $h$ above the top of the confining bars {[}Fig.
\ref{fig:wxwya2}(a)]. Here $h$ was determined by the height of the
laser sheet used to illuminated the particles and has an uncertainty
$\pm0.2$ mm. As $d$ increases $h$ decreases, indicating that the
sheath edge is moving into the rectangular depression.

Rather than taking $\omega_{x}$ and $\omega_{y}$ for each $d$ and
directly calculating $\alpha^{2}$, we fitted a line to $\omega_{x}$
and a quadratic curve to $\omega_{y}$ for $d>20$ mm, which spans
the zigzag and elliptical transitions. The average value of $\omega_{x}$
over this range is 7.47 rad/s. By dividing the two fitted curves we
determined $\alpha^{2}$ as a function of $d$ {[}Fig. \ref{fig:wxwya2}(b)].
We cover a wide range of anisotropies, where $\alpha^{2}$ decreases
with increasing $d$ from $\alpha^{2}\approx14$ to 2.5. If we could
increase $d$ to 50.8 mm to achieve an isotropic well, we would expect
$\alpha^{2}=1$. The slight increase of $\omega_{x}$ with $d$ implies
that $\kappa\propto\omega_{x}^{-2/3}$ increases weakly with $\alpha^{2}$.
If the only change in $\kappa$ is due to $\omega_{x}$, then we estimate
that $\kappa$ increases by $\approx6\%$ over the range of $\alpha^{2}$
values considered. 

\begin{figure}
\includegraphics[width=3.25in]{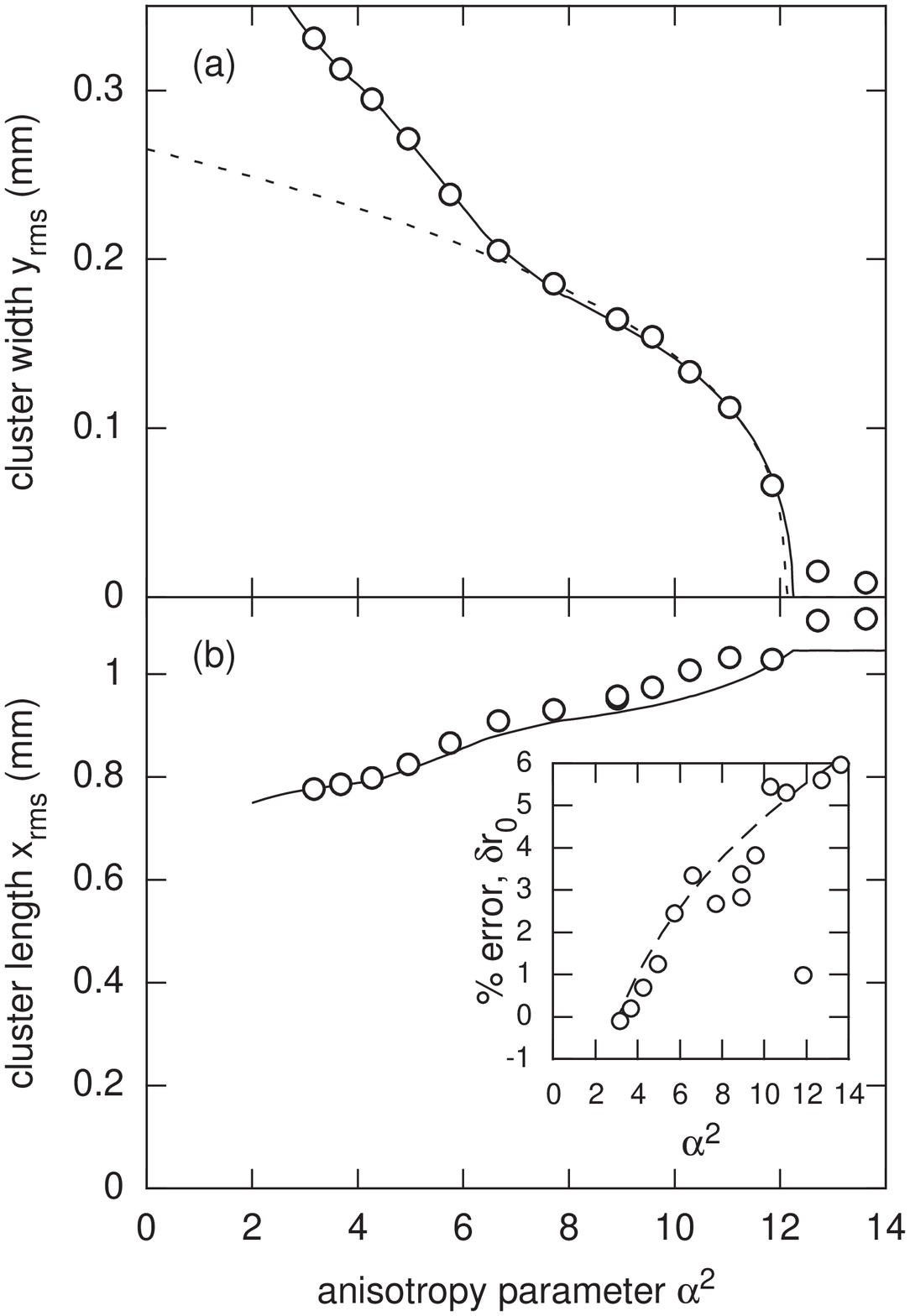}

\caption{\label{fig:width_vs_a2}(a) Measured cluster width $y_{rms}$ (open
circles), power law fit to data near the zigzag transition (dashed
line) and widths computed from the model for $\kappa=2.4$ and $r_{0}=1.25$
mm (solid line) vs the measured anisotropy parameter $\alpha^{2}$.
(b) Measured cluster length $x_{rms}$ (open circles) and lengths
computed from the model (solid line) vs $\alpha^{2}$. (Inset) Comparison
of the percentage difference between the $x_{rms}$ data and the model
(open circles) with the percentage change in $r_{0}\sim\omega_{x}^{-2/3}$
(broken line) vs $\alpha^{2}$. }

\end{figure}

The measured cluster width $y_{rms}$ and length $x_{rms}$ are plotted
vs $\alpha^{2}$ in Figs. \ref{fig:width_vs_a2}(a) and (b), respectively.
From the $y_{rms}$ data we observe a transition from the 1D configuration
to a 2D zigzag configuration at $\alpha^{2}\approx12$. From $\alpha^{2}\approx12$
to 6.5, the $y_{rms}$ data is well fitted by a power law with a critical
value $\alpha_{c}^{2}=12.1$ and an exponent of 0.35. The model solution
shows the same power law behavior \cite{tes8}. The divergence from
the power law fit at $\alpha^{2}\lesssim6.5$ indicates a second transition
\cite{tes8}, from the zigzag configuration to an elliptical configuration.
The elliptical structure \cite{tes4} occurs when the second (fifth)
particle in the cluster moves between the first and third (fourth
and sixth) particles, thereby moving onto the convex hull of the cluster.
Since the equilibrium shell structure for six particles in an isotropic
well is (1, 5), a further symmetry breaking transition is needed to
reach the $\alpha^{2}=1$ configuration.

For the first value of $d$ after the zigzag transition ($\alpha^{2}=11.8$),
the cluster was seen to flip between the two anti-symmetric zigzag
configurations. (Since we can vary $\alpha^{2}$ continuously it is
possible to tune for this behavior.) A zigzag configuration has two
degenerate states since the energy is invariant under a reflection
of particle coordinates $y\rightarrow-y$. In a stable zigzag configuration
these two states are separated by a potential barrier corresponding
to the higher energy of the unstable straight line configuration and
the zigzag state is bistable. Just above the zigzag transition the
barrier between the two states is low, and thermal fluctuations or
small periodic perturbations can cause the system to flip between
the two zigzag states, as was observed. For the next lower value of
$\alpha^{2}$ flipping was not seen, indicating that the potential
barrier between the two degenerate zigzag states was then too high.

Theoretical configurations which can be compared to the experiment
are found by minimizing Eq. (\ref{eq:UU0}), allowing us to predict
$x_{rms}=r_{0}\xi_{rms}$ and $y_{rms}=r_{0}\eta_{rms}$ as a function
of $\alpha^{2}$. To calculate a configuration three parameters, $\alpha^{2}$,
$\kappa$ and $n$, are required. We know $n$, and $\alpha^{2}$
has been measured, so we only need to determine the shielding parameter
$\kappa$. To do this we assume that $\kappa$ is independent of $\alpha^{2}$
and then minimize the sum of the squared differences between the measured
and computed values of $y_{rms}$, giving $\kappa=2.4$ and $r_{0}=1.25\;{\rm mm}$.
The Debye length $\lambda_{D}=r_{0}/\kappa=0.52\;{\rm mm}$ is comparable
to the inter-particle distance and the average charge $q=-1.2\times10^{4}e$.
These values are consistent with those found in previous experiments
for similar plasma conditions \cite{tes2,tes3,tes4,tes6,tes8}.

As shown in Fig. \ref{fig:width_vs_a2}(a), the cluster width $y_{rms}$
calculated using the model for $\kappa=2.4$ and $r_{0}=1.25$ mm
exhibits excellent agreement with the experimental data, including
reproducing the transition between the zigzag and elliptical configurations.
The agreement between the measured and predicted values of $x_{rms}$
provides a cross check on the model. We find that the predicted values
of $x_{rms}$ {[}Fig. \ref{fig:width_vs_a2}(b)] show good agreement
for $\alpha^{2}\lesssim4$, but systematically underestimate the measured
cluster length as $\alpha^{2}$ increases. For the longest clusters,
which are in a straight line configuration, the predicted cluster
lengths are about 6\% below the measured values. As shown in the Fig.
\ref{fig:width_vs_a2}(inset), this difference can be attributed to
an experimental increase in $r_{0}\sim\omega_{x}^{-2/3}$ due to the
measured decrease of $\omega_{x}$ with $\alpha^{2}$. This implies
that the average particle charge $q$ is effectively constant, as
previously observed for constant neutral pressure \cite{tes6,tes3}.
This increase in $r_{0}$ with $\alpha^{2}$ is not as apparent for
$y_{rms}$ since $y_{rms}\lesssim0.2x_{rms}$ and $y_{rms}\rightarrow0$
as $\alpha^{2}$ increases.

In summary, we have provided direct evidence that the width of a Yukawa
cluster exhibits power law behavior for the 1D to 2D zigzag transition
caused by decreasing the confining well anisotropy parameter $\alpha^{2}$,
confirming a previous prediction \cite{tes8}. Experiments were performed
using a dusty plasma with $n=6$ particles confined in the biharmonic
well above a rectangular depression. The width $d$ of the rectangular
depression was increased while the plasma remained on to decrease
$\alpha^{2}$ while the Debye shielding parameter $\kappa$ remained
essentially constant. The dependence of $\alpha^{2}$ on $d$ was
accurately determined by measuring the c.m. frequencies of the dusty
plasma. A transition from the zigzag configuration to an elliptical
configuration was also observed. The cluster width was found to be
in excellent agreement with the predictions of a model which assumes
identical particles confined in a 2D biharmonic well and interacting
through a Yukawa potential. From the fit to the model we found the
Debye length is comparable to the inter-particle distance, so that
Debye shielding significantly effects the physics of these clusters. 

\begin{acknowledgments}
Portions of this paper are taken from A. L. M.'s undergraduate physics
thesis.
\end{acknowledgments}

\end{document}